\documentclass[twoside]{article}
\usepackage{qic,epsfig}
\usepackage{amsmath}
\newcommand\numberthis{\addtocounter{equation}{1}\tag{\theequation}}
\usepackage{float}
\usepackage{polski}
\newcommand{\bra}[1]{\ensuremath{\left\langle#1\right|}}
\newcommand{\ket}[1]{\ensuremath{\left|#1\right\rangle}}

\begin{document}
\setlength{\textheight}{8.0truein}    %FOR 2ND PAGE ONWARDS=

%\runninghead{State-Independent Quantum Key Distribution with Two-Way Classical Communication $\ldots$}
           % {Radha Pyari Sandhir $\ldots$}

\normalsize\textlineskip
\thispagestyle{empty}
\setcounter{page}{1}

%\copyrightheading{Vol.}{No.}{Year}{Page Nos.}
%\copyrightheading{0}{0}{2003}{000--000}

\vspace*{0.88truein}

\alphfootnote

\fpage{1}

\centerline{\bf
%%%%%%%%%%%%%%%%%%%%%
%Put in titiles here
%%%%%%%%%%%%%%%%%%%%%
State-Independent Quantum Key Distribution with Two-Way Classical Communication}
%\vspace*{0.035truein}
%\centerline{\bf FOR QUANTUM INFORMATION AND COMPUTATION\footnote{Typeset the
%title in 10 pt Times Roman, uppercase and boldface.}}
\vspace*{0.37truein}
\centerline{\footnotesize
%%%%%%%%%%%%%%%%%%%%%%%%%%%%%%%%%%%%
%put authors' name and address here
%%%%%%%%%%%%%%%%%%%%%%%%%%%%%%%%%%%%
RADHA PYARI SANDHIR}
\vspace*{0.015truein}
\centerline{\footnotesize\it Department of Physics and Computer Science, Dayalbagh Educational Institute}
\baselineskip=10pt
\centerline{\footnotesize\it Uttar Pradesh 282005,
India}

\vspace*{0.225truein}
%\publisher{(received date)}{(revised date)}

\vspace*{0.21truein}

%% \abstracts{first paragraph}{second paragraph}{third paragraph}
%% If there is only one paragraph, just keep the second and third empty 
%% like the following one 
\abstracts{
A quantum key distribution protocol is proposed that is a variation of BB84 that provides raw key generation from correlations that violate a Bell-type inequality for single qubit systems and not entangled pairs. Additionally, it 1) is state-independent, 2) involves two-way classical communication, and 3) does not require basis matching between the two parties. The Brukner-Taylor-Cheung-Vedral (BTCV) time-like form of the Bell-CHSH inequality \cite{Bruk04,Tayl04} is employed as an eavesdropping check; sequential measurements lead to an inequality identical in form to the Bell-CHSH inequality, which relies only on the measurements performed with no regard for the qubit states. We show that this form manifests naturally from the non-commutativity of observables. 
}{}{}

\vspace*{10pt}

\vspace*{3pt}
%\communicate{to be filled by the Editorial}

\vspace*{1pt}\textlineskip    %) USE THIS MEASUREMENT WHEN THERE IS
   %) A SECTION HEADING
%\vspace*{-0.5pt}
%\noindent
%%%%%%%%%%%%%%%%%%%%%%%%%%%%%%%%
%put the text of the paper here
%%%%%%%%%%%%%%%%%%%%%%%%%%%%%%%%
\section{Introduction}\label{sect:intro}

Quantum key distribution (QKD) is a powerful cryptographic tool in quantum information theory that relies on quantum features such as superposition and non-locality to provide a secure key generation technique at a distance without any assumptions made on the computational power of a potential disruptor or eavesdropper. The structure of a generic QKD protocol consists of two components: 1) a quantum component that makes use of quantum systems and a quantum channel that needn't be secure to establish a raw key between two parties, and 2) a classical component that involves distillation of a key from the raw key through classical information reconciliation. A QKD protocol's purpose is to output a key, $K$, on one end, and an estimate of that key, $K'$, on the other.

Extensive strides have been made in this field, kicked off by the seminal BB84 protocol \cite{Benn84}, a prepare-and-measure protocol in which the first party prepares qubit states and transmits them to the second party, after which measurements are performed. Soon after, the seminal E91 protocol was proposed \cite{Ekert91}, involving an EPR scheme that employs the Bell-CHSH inequality \cite{CHSHOrig} to test for eavesdropping through the use of entangled qubit pairs that are shared between the two communicating parties, and subsequently using them for raw key generation. This operational employment of non-locality \emph{viz.} the Bell inequality \cite{BellInequalities} carved a path for intuitive device-independence, making redundant the requirement that the devices are characterized in detail or even trusted \cite{Masa11,Scar13}.

Thus far, in variations of the two seminal protocols, once the absence of an eavesdropper is established, the raw key generation relies on only those bits for which the measurement basis matches for both parties, thus leading to discardment on both ends. For instance, if qubit states are prepared as the polarization degrees of freedom of photons, only those bits obtained after measurements for which analyzer orientations matched on both ends are retained. It is interesting to note that even a BB84 scheme employing EPR pairs without a Bell inequality \cite{Benn92}, proposed soon after E91, an even larger amount of discardment occurs than the traditional BB84 scenario, due to the wider range of measurements both parties may randomly select.

Security of these protocols has extensively been investigated. BB84 has been proven secure for trusted devices with the assumption that quantum physics holds \cite{Shor00}, and due to the uncertainty principle \cite{Koashi06}. It has also been shown that a key distilled from correlations violating a Bell-type inequality is secure against a supraquantum eavesdropper limited only by the no-signalling condition \cite{Barr05,Acin06,Scar06}. Such protocols rely on entangled qubit pairs being shared with both parties beforehand. Indeed it is the monogamy of spatial correlations that ensures secure communication between parties \cite{Pawl10}. Correlations revealed by sequential measurements are polygamous, or monogamous in a weaker sense \cite{Nowa17, Shen17}. Yet, an augmentation with a time-like version of the Bell inequality protects against a weakness BB84 appears to have \cite{Shen17}:  susceptibility when an eavesdropper introduces higher dimensional ``cheat states".

At the heart of these secure protocols and variations thereof lie the generation and manipulation of quantum states. The preparation of quantum states is in itself a vast area of study, prone to obstacles such as decoherence \cite{Fuji00}.  While the ideal experimental scenario involves the preparation of a quantum system in a pure state, the reality is that the state inevitably interacts with its environment to evolve into more of a statistical mixture of pure states, or a mixed state. In this manner, quantum information is irreversibly leaked. While advances in quantum optics hope to curb this leakage (see \cite{Sera12} and citations therein), applications that are state-independent would eliminate the need for complex processes to prepare and manipulate quantum states.

Apart from the matters of security and quantum state preparation, increasing the key rate is an important goal. The key rate is the ratio of the length of the final secure key to the length of the raw key. Making use of two-way classical communication during the information reconciliation stage has been shown to not only be secure \cite{Gotte03, Kraus07}, but to also increase this key rate \cite{Wata07}. 

The protocol proposed in this paper is a variation of BB84 that establishes state-independence, and makes use of a violation of a Bell-type inequality, as well as two-way classical communication. It has a number of features, taking into account the strides outlined above. The features include:

\begin{itemize}
\item State-independence
\item Two-way classical communication
\item Raw key generation from correlations that violate a Bell-type inequality for single qubit systems and not entangled pairs
\item No requirement of measurement matching in order to retain data
\end{itemize}

This state-independent quantum key distribution protocol with two-way classical communication (SIQKD2CC) has at its core measurements that are not space-like separated, but time-like separated. To realize this, we employ the Brukner-Taylor-Cheung-Vedral (BTCV) formalism \cite{Bruk04,Tayl04}, which makes use of sequential projective measurements on a single system, as opposed to measurements on space-like separated coupled systems.

The BTCV formulation relies solely on the measurement operators, and one can therefore not make a comment on the non-classicality of the state. However, in some sense, one can say that all states can violate this inequality, even fully mixed ones \cite{Fedr11}. Correlations in time can be considered manifestations of sequential measurements. BTCV provides an operational probabilistic framework that can be applied to enhance probabilities of success in classical protocols such as computational complexity tasks, demonstrated in \cite{Bruk04,Tayl04}. In the SIQKD2CC protocol, the BTCV formalism provides a probabilistic tool that allows two parties to detect an eavesdropper even if the initial state is completely mixed -- thus rendering it truly state-independent. The current text focuses on BTCV and thus temporal correlations as a potential resource for state-independent protocols, though a discussion on security has been provided in Section~3.1.1.

We begin with a brief overview of the time-like Bell-CHSH inequality in Section~\ref{sect:BTCV} with important notes on the origin of the state-independence and whether the correlations can be considered non-classical. We show that the form of this inequality arises naturally from non-commutativity of observables. We then proceed to the SIQKD2CC protocol in Section~\ref{sect:SIQKD}. 

\section{BTCV Time-like Bell-CHSH Inequality}\label{sect:BTCV}

We briefly discuss the time-like Bell-CHSH inequality as given in \cite{Bruk04,Tayl04}, further studied in \cite{Fritz10}, and experimentally verified in \cite{Fedr11, Ring18}. Note that the Leggett-Garg (LG) inequality \cite{Legg85, Emary14} is also a time-like inequality similar in form to Bell-CHSH, but the BTCV inequality provides an operational formalism that is applicable to the context of this work and will therefore be the focus of the time-like scenario discussed here.

We start with the assumptions. The original Bell-CHSH inequality \cite{BellInequalities,CHSHOrig} arises from classical assumptions made on two space-like separated parties: i) realism, that each party has definite properties prior to measurement, and ii) locality, that neither party can be influenced by measurements made on the other. In the time-like scenario, realism can still be considered a non-controversial assumption, yet the notion of `locality in time' differs from locality in the traditional sense, as measurements are performed on a single system.  Both BTCV and LG inequalities assume `non-invasiveness', a notion originally theorised by Leggett and Garg, in which a measurement cannot influence a system at an earlier or later time, yet, quantum physics is, at its heart, an invasive theory.

There is, therefore, some ambiguity about the implications of such assumptions reflecting genuine entanglement in time, or more generally: non-classicality of temporal correlations in the context of sequential measurements. Moreover, understanding true `non-locality' in time remains to be explored. However, we show in Section~\ref{sect:proof} that this form manifests naturally from non-commutating observables, and so, a comment on entanglement in time is not required. Thus, our assumptions lie in non-commutativity of observables in quantum physics, which naturally provides a form equivalent to the assumptions of local realism in time. It is this alternative quantum origin that we focus on, as opposed to the assumptions made in \cite{Bruk04,Tayl04}.

The BTCV inequality is constructed as follows. Consider an arbitrary two-level mixed state: $\rho=\frac{1}{2}[I+\vec{\sigma}\cdot\vec{r}]$. Probability of measurement with observable $A=\vec{\sigma}\cdot\vec{a}$ yielding outcome $k$ at time $t_1$ is $Tr[\rho\Pi_{A}^k]$, where $\Pi_{A}^k$ is the projector associated with outcome $k$. Probability of measurement with observable $B=\vec{\sigma}\cdot\vec{b}$ yielding outcome $l$ at time $t_2$ given that the outcome $k$ was obtained at time $t_1$ is $Tr[\Pi_{A}^k\Pi_{B}^l]$, where $\Pi_{B}^l$ is the projector associated with outcome $l$. Therefore, the correlation function between the measurements at the two times is given by:

\begin{equation}
C(\vec{a},\vec{b})=\sum_{k,l=\pm 1}klTr[\rho\Pi_{A}^k]Tr[\Pi_{A}^k\Pi_{B}^l]
\label{eq:timelikecorr}
\end{equation}

where $\Pi_A^k=\frac{1}{2}[I+k\vec{\sigma}\cdot\vec{a}]$, $\Pi_B^l=\frac{1}{2}[I+l\vec{\sigma}\cdot\vec{b}]$. 

This leads to: 

\begin{equation}
C(\vec{a},\vec{b})=\vec{a}\cdot\vec{b}, 
\end{equation}

This correlation function is similar in form to the quantum mechanical correlation function for the spin-$\frac{1}{2}$ singlet state, as solely being dependent on the angle between $\vec{a}$ and $\vec{b}$. Moreover, it is remarkably state-independent, that is, it is independent of the initial state $\rho$. Thus, whether the initial qubit state was mixed or pure is irrelevant, as once the initial measurement is performed, the resulting state is that which corresponds to the projector associated with the outcome.

Note that any qubit dynamics between $t_1$ and $t_2$ can also be represented by a rotation matrix $R\in SO(3)$. Prior to the second measurement, the state is $R(k\vec{a})=kR(\vec{a})$, where $k=\pm 1$. The probability for obtaining the outcome $l$ is $\frac{1}{2}(1+kl\vec{b}\cdot R(\vec {a}))$. The correlator becomes $C=R(\vec{a})\cdot \vec{b}$ \cite{Fritz10}.

Through classical constraints, the bound imposed on the time-like analog of the Bell-CHSH expression is:

\begin{equation}\label{eq:timelikebellchsh}
\mathcal{B}\equiv | C(\vec{a_1},\vec{b_1})+ C(\vec{a_1},\vec{b_2})+ C(\vec{a_2},\vec{b_1})-C(\vec{a_2},\vec{b_2})|\leq 2
\end{equation}

Maximal violation occurs at $2\sqrt{2}$ for $\vec{a_1}=\frac{1}{\sqrt{2}}(\vec{b_1}+\vec{b_2})$ and $\vec{a_2}=\frac{1}{\sqrt{2}}(\vec{b_1}-\vec{b_2})$. Therefore, the Tsirelson bound \cite{Tsirelson80} can be reached by appropriate measurements on the Bloch sphere.

\subsection{Measurement at $t_2$ Where $t_1<t_2<t_3$}

If three sequential measurements $(\vec{\sigma}\cdot\vec{a}),(\vec{\sigma}\cdot\vec{b}),(\vec{\sigma}\cdot\vec{c}$) are considered instead of two, at times $t_1<t_2<t_3$, then the correlation function between $t_1$ and $t_3$ is \cite{Bruk04,Tayl04}:

\begin{align}\label{eq:3measurements}
C(\vec{a},\vec{c})&=\sum_{k,l,s=\pm 1}klsTr[\rho\Pi_{A}^k]Tr[\Pi_{A}^k\Pi_{B}^l]Tr[\Pi_{B}^l\Pi_{C}^s]\\
&=(\vec{a}\cdot\vec{b})(\vec{b}\cdot\vec{c}) 
\end{align}

Therefore: 1) the quantum correlation function for measurements at $m$ instances can be considered a product of two-fold temporal correlations, and 2) measurement at $t_2$ disrupts correlations between measurements at $t_1$ and $t_3$ as the time-like Bell-CHSH inequality between $t_1$ and $t_3$ is no longer violated - a straightforward algebraic exercise.

\subsection{Non-commutativity and the Time-like Bell-CHSH Inequality}\label{sect:proof}

For the sake of conceptual clarity, we show how the Bell-CHSH inequality derived by \cite{Bruk04,Tayl04} can be understood as a natural arisal from the non-commutativity of observables and not necessarily as a reflection of non-classical correlations in time. Indeed, the non-classicality in this context can be attributed to non-commutativity and the symmetric nature of conditional probabilities in quantum physics.

For this purpose, we take into consideration the \emph{pseudo-projection} formalism as given in \cite{Adhi18}, which provides an operational framework based on the notion of a pseudo-projection -- which is, in essence, the quantum representative of the classical indicator function of a probabilistic joint event. This framework addresses the question originally posed by Fine \cite{Fine82, Fine82_2}: is the construction of joint probabilities possible for a quantum state? The test indicates whether a joint probability distribution can be constructed for a state, which would imply that there exists a classical system that can mimic this state with respect to the joint probability distribution. In this regard, the state can be considered `classical'. The pseudo-projection formalism gives rise to negative operators and thus non-classical probabilities that arise from an operationally Boolean framework.

The representative of the classical indicator function in quantum physics is the projection operator, $\Pi$. Consider a quantum state: a density operator, $\rho$, in a Hilbert space, $\mathcal{H}$. Obtaining the probability for observable $A$, acting on $\rho$, resulting in outcome $a_i$, is equivalent to obtaining the extent of overlap between projection operator $\Pi_{a_i}=\ket{a_i}\bra{a_i}$ and the state, namely: $P(A=a_i)=Tr[\Pi_{a_i}\rho]$.  

When considering a joint event involving $\Pi_{a_1}$ and $\Pi_{a_2}$, the mapping from the classical domain to the quantum domain is inconsistent due to the uncertainty principle. The quantum representative considered in the pseudo probability formalism is the symmetrized product that follows Weyl ordering \cite{Weyl27}: $\Pi_{a_1a_2}^{A_1A_2}=\frac{1}{2}\{\Pi_{a_1}^{A_1}\Pi_{a_2}^{A_2}\}$. The ensuing probabilities that arise from this pseudo-projection operator are called `pseudo-probabilities', and contain information on the classical (or lack thereof) nature of the system. Namely, any state that manifests pseudo-probabilities that lie beyond the expected bound $[0,1]$ can be considered to be non-classical.

In order to establish an inequality equivalent in form to the state-independent Bell-CHSH inequality, we impose assumptions on the probabilities such that they are non-negative. Thus, a violation of such an inequality would indicate non-classicality.

\vspace*{12pt}
\noindent
{\bf Proposition:} A state-independent Bell-CHSH inequality can be derived by non-commuting observables, provided the first observable, $A$, shares an eigen basis with the original state, $\rho$, that is: $[A,\rho]=0$.

\vspace*{12pt}
\noindent
{\bf Proof:} Consider three non-commuting dichotomic observables $A$, $B$ and $C$, where $A$ shares an eigen basis with state $\rho=\frac{1}{2}[1+\vec{\sigma}\cdot\vec{p}]$. Thus, the projectors of $A$ corresponding to outcomes $\pm 1$ are:
\begin{equation}
\Pi_{A\pm}=\frac{1}{2}[1\pm\vec{\sigma}\cdot\vec{p}]
\end{equation}

Projectors of $B$ and $C$ take a similar form, defined by variables $\vec{b},\vec{c}$ respectively:

\begin{align*}
\Pi_{B\pm}=\frac{1}{2}[1\pm\vec{\sigma}\cdot\vec{b}]\\
\Pi_{C\pm}=\frac{1}{2}[1\pm\vec{\sigma}\cdot\vec{c}]\numberthis
\end{align*}

As $A,B,C$ are non-commuting, $Tr[\Pi_{i\pm}\Pi_{j\pm}]\geq0$ always, where $i,j=A,B,C$ and $i\neq j$. 

Without loss of generality, we consider the symmetrized product of projectors $\Pi_{A+},\Pi_{B+},\Pi_{C+}$ in the form of a pseudo-projection operator\footnote{Note that the assumption of probabilities being non-negative is what gives rise to the classical bound of the inequality. These probabilities are different from quantum probabilities arising from measurements in, say, a Bell-type experiment.}, and omit the outcome index $+$ for convenience:

\begin{align*}
\Pi_{ABC}&=\frac{1}{16}\Big[\Pi_A\Pi_B\Pi_C+\Pi_C\Pi_B\Pi_A\Big]\\
&=\frac{1}{16}\Big[2+2(\vec{\sigma}\cdot\vec{p})+2(\vec{\sigma}\cdot\vec{b})+2(\vec{\sigma}\cdot\vec{c})+2(\vec{\sigma}\cdot\vec{p})\big((\vec{\sigma}\cdot\vec{b})+(\vec{\sigma}\cdot\vec{c})\big)+2(\vec{\sigma}\cdot\vec{b})(\vec{\sigma}\cdot\vec{c})\\
&+(\vec{\sigma}\cdot\vec{p})(\vec{\sigma}\cdot\vec{b})(\vec{\sigma}\cdot\vec{c})+(\vec{\sigma}\cdot\vec{c})(\vec{\sigma}\cdot\vec{b})(\vec{\sigma}\cdot\vec{p})\Big]\numberthis
\end{align*}

This pseudo-projection operator is Hermitian, but not indempotent, ergo `pseudo'. Taking the trace, one finds:

\begin{align*}
Tr[\Pi_{ABC}]&=\frac{1}{16}\Big(4+4\vec{p}\cdot(\vec{b}+\vec{c})+4(\vec{b}\cdot\vec{c})\Big)\numberthis
\end{align*}

Now, consider the joint measurements of $B$ and $C$ on state $\rho$. Classically, the probability would be positive if not zero, and thus $Tr[\Pi_{ABC}]\geq 0$, that is, $1+\vec{p}\cdot(\vec{b}+\vec{c})+(\vec{b}\cdot\vec{c})\geq 0 $, as $A$ shares an eigen basis with $\rho$. Taking $\vec{b}\rightarrow -\vec{b}$ and $\vec{c}\rightarrow -\vec{c}$, and combining the two we find:

\begin{align*}
1+(\vec{b}\cdot\vec{c})\geq 0 \numberthis
\label{eq:probbc}
\end{align*}

Note that Eq.~\ref{eq:probbc} is the symmetric conditional probability employed in the time-like correlation function given in Eq.~\ref{eq:timelikecorr}, namely:

\begin{align*}
P(B|C)=P(C|B)&=Tr[\Pi_B\Pi_C]\\
&=Tr\bigg[\frac{1}{2}(1+\vec{\sigma}\cdot\vec{b})\frac{1}{2}(1+\vec{\sigma}\cdot\vec{c})\bigg]\numberthis
\end{align*}

This symmetrization is at the heart of quantum physics, being impossible in the classical scenario, and is a feature that is exploited by the Bell-CHSH inequality.

In order to bring this to the Bell-CHSH form, we consider four orientations $\vec{b_1},\vec{b_2},\vec{c_1},\vec{c_2}$. Per the classical law of total probability and the Kolmogorov definition of conditional probability, $\sum_k P(C_k|B)=1$, and thus we may impose the classical bound:

\begin{align*}
&P(C_1|B_1) + P(C_2|B_1) + P(C_1|B_2)  - P(C_2|B_2) \geq 0\\
&\implies 1+(\vec{b_1}\cdot\vec{c_1})+1+(\vec{b_2}\cdot\vec{c_1})+1+(\vec{b_1}\cdot\vec{c_2})-1-(\vec{b_2}\cdot\vec{c_2})\geq 0\\
&\implies \vec{b_1}\cdot(\vec{c_1}+\vec{c_2})+\vec{b_2}\cdot(\vec{c_1}-\vec{c_2})\leq 2 
~~\square\ \numberthis
\end{align*}

\subsection{Short Note: Non-classicality and Temporal Correlations}\label{sect:shortnote}

That the time-like Bell-CHSH inequality of the BTCV prescription can be violated maximally is on its own not enough to suggest that the correlations are non-classical, even though the inequality arises from a non-classical construction as previously mentioned. The Toner-Bacon (T-B) 1-bit protocol \cite{Tone03}, used to simulate the singlet state, can easily be implemented to simulate temporal correlations \cite{Brier15}. Moreover, a sequence of T-B protocols can be implemented to mimic sequential projective measurements, as the correlations are factored into dot product pairs as shown in Eq.~\ref{eq:3measurements}. 

An assessment of non-classicality, defined by the classical simulability of a temporal correlation, is given in  \cite{Brier15}. A temporal correlation function of an $d$-level physical system is said to be non-classical if every classical algorithm that simulates the function requires more than $\log_2 d$ bits of classical communication at some point during the simulation. In \cite{Brier15}, a lower bound is provided for the number of sequential measurements to be performed in order for the correlations to be non-classical. For a qu$d$it, the number of measurements, $n$, should follow $n\geq d$. 

\begin{figure} [H]
\centerline{\epsfig{file=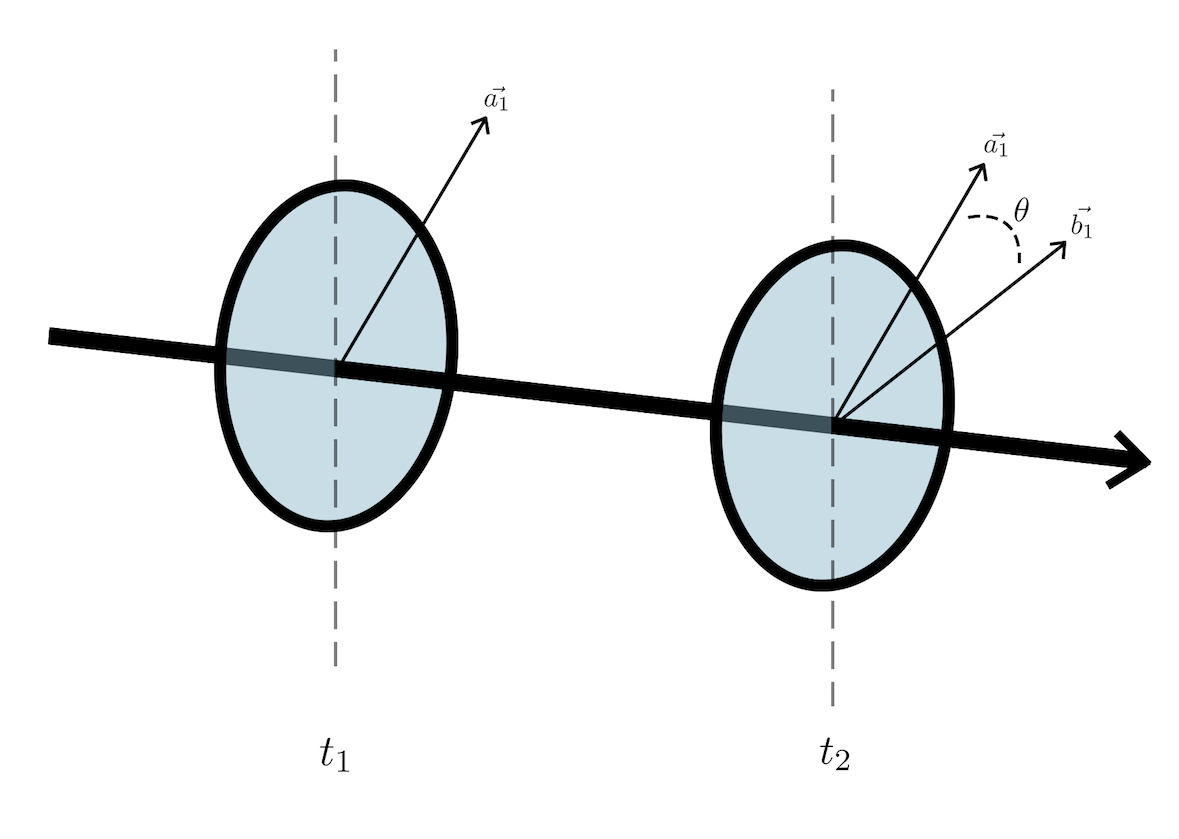, width=11.2cm}}
\vspace*{13pt}
\fcaption{\label{fig:qkdoptics}Representation of an experimental realization of the correlation function $C(\vec{a_1},\vec{b_1})=\cos\theta$}
\end{figure}

This is important to note before making a claim on the non-classicality of the correlations. However, in the SIQKD2CC protocol proposed ahead, it is unnecessary to ensure non-classical correlations in the sense described above, as the protocol relies instead on a violation of the time-like Bell-CHSH inequality, which is an algebraic consequence of the non-classical conditional probabilities. Indeed, this sequence of measurements may be realized through classically polarized electromagnetic waves, a representation of which is shown in Figure~\ref{fig:qkdoptics}, considering the use of two half wave plates with principle axes coinciding with $\vec{a_1}$ and $\vec{b_1}$ respectively. The correlation function $C(\vec{a_1},\vec{b_1})=\vec{a_1}\cdot\vec{b_1}$ relies on the angle between the two.

Because this sequence of measurements leads to an inequality identical in form to the Bell-CHSH inequality, we continue to make use of this terminology. With this disclaimer, we proceed to the protocol.

\section{State-independent Secure Quantum Key Distribution with Two-Way Classical Communication}\label{sect:SIQKD}

\subsection{The Structure of a QKD Protocol and Motivations Behind SIQKD2CC}\label{sect:discussion}

\begin{figure} [H]

\centerline{\epsfig{file=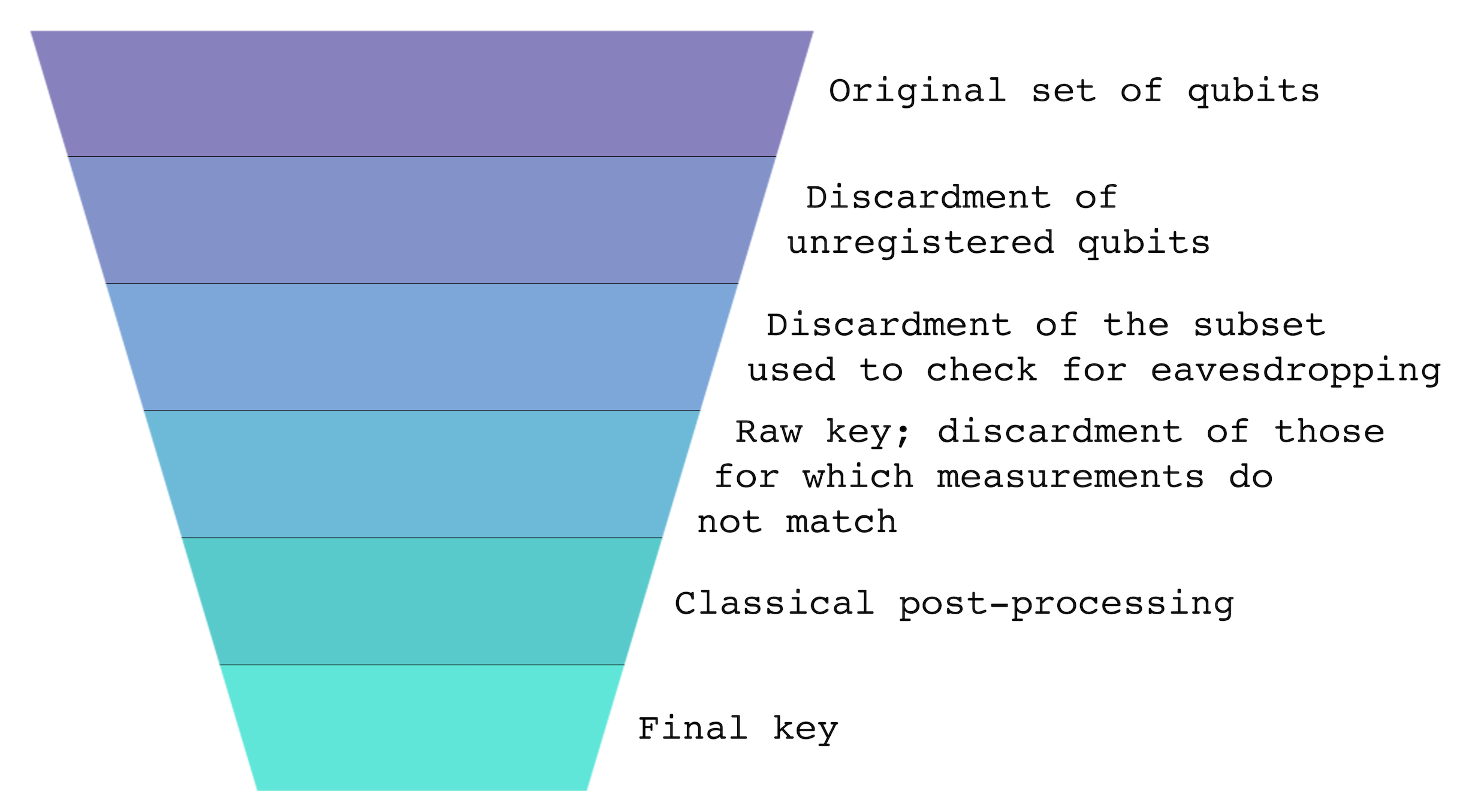, width=12.2cm}} 
\vspace*{13pt}
\fcaption{\label{fig:frustum} The structure of a generic QKD protocol in light of the units of information that are processed.}
\end{figure}

The generic QKD protocol, tracing back to the original BB84, can be thought of as the distillation of a key from a sequence of qubit states shared between two parties, Alice and Bob. We consider the sequence to be a sequence of units of information, and the structure of a generic QKD protocol can be understood by looking at what happens to these units. Through the process, more and more units of information are discarded. One can visualize the structure as an inverted frustum, as illustrated in Figure~\ref{fig:frustum}. 

At the top we have a complete sequence of qubits, such as a sequence of states prepared by Alice and transmitted over a quantum channel to Bob in the light of BB84 and its derivatives, or a set of entangled qubit pairs shared by the two, such as the E91 protocol and its derivatives. 

Qubits that fail to register are discarded. A subset of the sequence is used post measurement on both sides in order to detect the presence of an eavesdropper. The units of information consist of correlated classical data. Once the absence of a third party is established, that subset is discarded. Most protocols then require Alice and Bob to discard those units for which their measurements do not match. For instance, in a modification of the BB84 that is equivalent to E91 \cite{Benn92}, Alice and Bob can choose among a continuum of measurements in a range between axes $0^o$ and $90^o$. They discard those for which their measurements do not match, leading to a heavy reduction in units of information, as there are a large number of possible combinations they may choose from. 

The implementation of a successful quantum key distribution protocol requires two parts: 1) the quantum part, which establishes a raw key between two parties, and 2) classical post-processing, in which privacy amplification and information reconciliation lead to a distillation of the final one-time pad from the raw key. This distillation process further reduces the units of information at hand.

Thus, there is generally a large amount of `wastage', and it would be a prudent goal to reduce the amount of discardment that occurs wherever possible.

\subsubsection{A note on state-independence and security}

A BB84-like scenario can technically be rendered state-independent by performing Von Neumann measurements on arbitrary states. Yet, such protocols often require the first party, Alice, to encode classical strings, thus making state preparation an important part of the process. 

For this reason, preparation and manipulation of qubit states in QKD is an important task.

Therefore, not only is minimizing discardment such as that described above a prudent goal, targeting state-independence without compromising on security appears to be one as well. Moreover, in protocols that involve sharing of entangled pairs between the two parties beforehand, a worst-case assumption is that the eavesdropper prepares the states in the first place. It would be beneficial to avoid a scenario where this assumption is necessary.

The SIQKD2CC protocol proposed here is a variation of BB84 that generates a raw key through a state-independent method that makes use of the BTCV inequality  \cite{Bruk04,Tayl04}. In protocols involving spatial correlations, monogamy of entanglement ensures the condition of secure communication \cite{Pawl10}: $I(A:B) > I(A:E)$ \cite{Csis78}, where $I(A:B), I(A:E)$ are the mutual information between Alice and Bob, and Alice and Eve, respectively. Contrary to the spatial scenario, temporal correlations allow a simultaneous violation of pairwise BTCV inequalities, and are considered polygamous. However, these correlations are still bounded by a polygamy relation \cite{Ring18}: it is not possible to reach the algebraic maximal simultaneous violation for all pairs, and it has been shown that the BTCV inequality between the end parties is satisfied if Eve makes use of projective measurements. Additionally, monogamy has been shown to manifest when considering a modification to the consistent histories approach \cite{Nowa17}. 

When Bob's basis is selected to be $B_\pm=\frac{X\pm Y}{\sqrt{2}}$ to Alice's choice of $X$ and $Y$ -- made use of in the toy problem outlined in Section~\ref{sect:toyproblem} -- the scenario becomes equivalent to \cite{Shen17} without their requirement of basis matching, and which is secure against higher dimensional eavesdropping attacks. Note that \cite{Shen17} additionally proposes temporal monogamy related to the no-signalling condition, which is weaker than its spatial counterpart. A detailed proof of security in light of the monogamy/polygamy relations falls under the domain of future work, while the current text focuses on BTCV and thus temporal correlations as a potential resource for state-independent protocols.

Nonetheless, the classical post-processing implemented in the protocol outlined ahead crucially takes into consideration Eve somehow obtaining partial information of the final key. Classical pre-processing is conducted in the light of \cite{Wata07}, though prior to the quantum part as opposed to information reconciliation, and continues with two-way classical communication during the distillation stage. The protocol involves two separate parties, Alice and Bob. Alice locally randomly generates a binary string $X_2$ within her sealed lab, a part of which will be used to generate the final key. The crux of this protocol is to convey this string securely to Bob, without a third eavesdropping party, Eve, gaining knowledge of it. Alice outputs a key on one end, and Bob estimates that key. 

We begin with the labelling conventions and assumptions behind this protocol.

\subsection{Labelling Conventions}

Alice performs measurements with dichotomic observables $A_i$, which rely on parameters $\vec{a}_i$. Likewise, Bob's dichotomic observables are denoted by $B_i$ with parameters $\vec{b}_i$. 

Auxiliary random variables are made use of to aid with the execution of this protocol, drawing inspiration from secure prescriptions of privacy amplification and information reconciliation in \cite{Benn95,Barr05,Scar06,Acin06,Wata07}. Each is a string of $n$ bits, \emph{viz.} $X_1=\{x_{i1}\}_{i=1}^{i=n}$. We shall denote random variables in Alice's subsystem by $X_i$ and in Bob's subsystem as $Y_i$. Random variables $X_1$ and $Y_1$ are classically correlated with a probability distribution $P(X_1,Y_1)$, and are shared between Alice and Bob. Eavesdropping party, Eve, may have partial or complete information of the distribution $P(X_1, Y_1)$, which still allows for safe key distillation \cite{Benn95}. The string that will ultimately be used to determine the key, $X_2$, is locally randomly generated in Alice's sealed lab to guarantee Eve has no knowledge of it. 

The following additional variables are defined:
\begin{align*}
U_1=X_1\oplus X_2 \\
V_1=Y_1 \oplus Y_2 \numberthis
\end{align*}
in which $\oplus$ indicates bitwise binary addition, and where $Y_2$ will be defined at a later stage. Note: all `addition' is modulo 2 for bits, but can be extended for higher dimensions as well.

\subsection{The Assumptions}

\begin{itemlist}
\item What is referred to as a qubit is in actuality a qubit ensemble, for the determination of probabilities.
\item Qubits, that is, the qubit ensembles, are prepared in Alice's sealed laboratory.
\item Observables $A_1,A_2,B_1,B_2$ are selected such that maximal violation of the time-like Bell-CHSH inequality given by Eq.~\ref{eq:timelikebellchsh} can be attained. This may be selected by Alice and Bob prior to the execution of the protocol.
\item Qubit transmission happens over an error-free quantum channel.
\end{itemlist}

\subsection{The SIQKD2CC Algorithm}

\subsubsection{The Quantum Component}

\begin{figure} [H]

\centerline{\epsfig{file=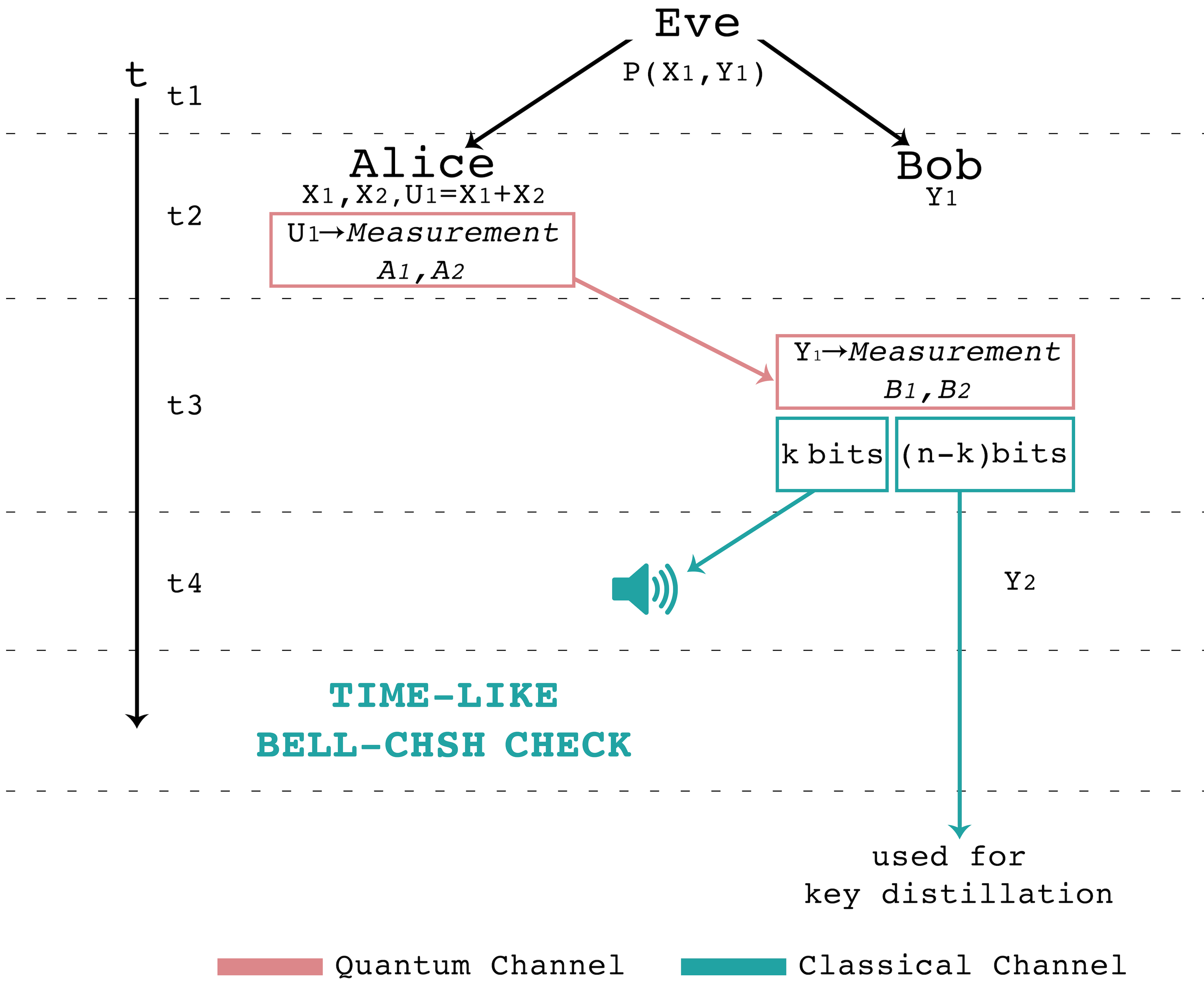, width=11.2cm}} 
\vspace*{13pt}
\fcaption{\label{fig:qkdtimealgo}The generation of a raw key during the quantum part of SIQKD2CC.}
\end{figure}

The generation of a raw key, that is, the quantum part of the algorithm, is illustrated in Figure~\ref{fig:qkdtimealgo}. Note, once again, that the use of `quantum' in this context does not imply the use of non-classical correlations nor necessarily quantum measurements, in the light of Section~\ref{sect:shortnote}.

The method is as follows:

\begin{romanlist}
\item At time $t_1$, Alice and Bob share classically correlated random variables $X_1$ and $Y_1$ with probability distribution $P(X_1,Y_1)$. Eve may even have partial or complete information about this distribution, and may have provided Alice and Bob with these classical strings.
\item Alice has at her disposal $n$ qubits, and at time $t_2$, performs measurements on them according to the binary string $U_1$. An example of such a scheme is the selection of observables: 0 - $A_1$, 1 - $A_2$.
\item The qubits are transmitted from Alice through the error-free quantum channel to Bob. 
\item At time $t_3$, Bob performs measurements on the first $k$ qubits, which will later be used for a time-like Bell-CHSH check. He selects his observables $B_1$ and $B_2$ based on string $Y_1$, in a manner similar to that of Alice selecting her observables.
\item At time $t_4$, Bob publicly discloses his observables and probabilities for the first $k$ qubits. Recall that rather than this referring to individual qubits, each is a qubit ensemble so that probabilities may be obtained.
\item Alice at $t_5$, having complete knowledge of the $k$ qubits measured at times $t_2$ and $t_3$, determines whether the time-like Bell-CHSH inequality has been violated. If Eve had accessed the quantum channel between $t_2$ and $t_3$ then correlations between measurements performed at those times would have been disrupted, and the Bell-CHSH inequality would be satisfied. If no eavesdropper is detected, the remaining $n-k$ qubits are used for key distillation. 
\item Bob performs measurements in accordance with $Y_1$ on the remaining $n-k$ qubits. Classical string $Y_2=\{y_{i2}\}_{i=k+1}^{i=n}$ is the result Bob obtains.  He calculates $V_1=Y_1\oplus Y_2$, where the first $k$ bits in $Y_1$ are disregarded. In a similar manner, the rest of the protocol makes use of only the last $n-k$ bits of each string. Thus, $X_2$ now denotes the last $n-k$ of the original bit string $X_2$.
\end{romanlist}

\subsubsection{The Classical Component}

The key is distilled with the help of two-way public discussion over a classical channel as follows:

\begin{itemlist}
\item Step 1: Alice sends Bob the classical information about $U_1$ by operating with a parity check matrix, $M$. That is, she sends the string $MU_1$, as this is more secure than Alice simply sending $U_1$  \cite{Wata07}. As a parity matrix, $M$ follows, for a linear code $U_1$, $u_1M^T=0$ for all $u_1\in U_1$.
\item Step 2: Bob determines $W_1=MU_1\oplus V_1$, and discloses it to Alice.
\item Step 3: Alice determines $U_2=MX_1\oplus W_1$ and discloses it to Bob.
\item Step 4: Bob predicts $X_2$ through $X'_2=U_2\oplus V_1$.
\item Step 5: Alice and Bob then use a pre-determined hash function $f$ to distill the key: $K_A=f(X_2)$, $K_B=f'(X'_2)$, where $f'$ is a function of $f$ and $M$. This way, Bob need not have knowledge of $M$.
\end{itemlist}

Table~\ref{tab:classprocessing} outlines the classical post-processing that occurs in each of the two subsystems.

\vspace*{4pt}   
\begin{table}[H]
\centering
\tcaption{SIQKD2CC Classical Post-processing in Each of the Two Subsystems}
\begin{tabular}{ |c|c| } 
 \hline
 Alice & Bob  \\ 
 \hline
$MU_1=M(X_1\oplus X_2)$ & $V_1=Y_1\oplus Y_2$ \\ 
$U_2=MU_1\oplus V_1\oplus MX_1$ & $W_1=MU_1\oplus V_1$\\ 
$-$ & $X'_2=U_2\oplus V_1=MX_2$ \\
 \hline
\end{tabular}
\label{tab:classprocessing}
\end{table}

\subsection{Toy Problem}\label{sect:toyproblem}

To demonstrate SIQKD2CC, a simple example is considered with $n=5, k=2$. Eve distributes $X_1$ and $Y_1$ to Alice and Bob. Assume the random variables share the trivial distribution in which bit sequence $\overline{Y_1}=X_1$. The binary string $X_2$ is locally randomly generated in Alice's sealed lab as 10110, of which only the last three bits, 110, will ultimately be used for key distillation. For the purpose of this example, assume parity matrix $M$ is the identity matrix. Assume the following measuring mechanism is implemented in their respective subsystems: $0\equiv A_1=Z, B_1=\frac{Z+X}{\sqrt{2}}, 1\equiv A_2=X, B_2=\frac{Z-X}{\sqrt{2}}$, which has been experimentally verified to maximally violate the Bell-CHSH inequality using quantum optics \cite{Fedr11}. This is equivalent to the selection of basis $\ket{H},\ket{V},\ket{D},\ket{A}$ (see \cite{Ali17} for implementation using optics).  Additionally, note that this choice of Bob's basis matches that of Bob's additional basis in \cite{Shen17}.

\vspace*{4pt}   
\begin{table}[H]
\centering
\tcaption{SIQKD2CC Toy Example}
\begin{tabular}{ |c|c| } 
 \hline
 Alice & Bob  \\ 
 \hline
 $X_1=10101$ & $Y_1=01010$ \\ 
  $X_2=10110$ & - \\ 
 $U_1=X_1\oplus X_2= 00011$& - \\
Observable sequence: & - \\
 $A_1,A_1,A_1,A_2,A_2$ & -\\
 - & Observable sequence for first $k=2$ bits: $B_1, B_2$ \\ 
 Bell-CHSH check & - \\
 - & $Y_2= 100$ (say) on remaining $n-k=3$ bits \\
 - & $V_1=Y_1\oplus Y_2=110$ \\
 - & $ W_1=MU_1\oplus V_1=101$ \\
 $U_2=MX_1\oplus W_1=000$ & - \\
 - & $X'_2=U_2\oplus V_1=110$\\
 $key_A=f(X_2)$ & $key_B=f'(X'_2)$\\
 \hline
\end{tabular}
\label{tab:siqkdtoyproblem}
\end{table}

Assuming that the hash function merely acts as the identity map, the ideal key length attainable is $n-k=3$ bits, and the ideal key rate attainable is $1$.

\subsection{Short Note: Key Rate and Hash Function}

Note that the key rate portrayed is only an ideal.  If $E$ is the random variable that summarizes Eve's information about $X_2$, and $P(E,X_2)$ is the joint distribution of the two, the length of the final distillable key $r$ and hence, the key rate, depends on the hash function selected during the final distillation stage of the protocol and the constraints that $P(E,{X_2})$ must satisfy. 

A hash function $f$ is a function that can be used to map an arbitrarily sized string to a fixed size string. The function $f: X_2 \rightarrow \{0,1\}^{r_f}$ is in general randomly selected from an appropriate class of maps, $F$, where $r_f$ is the length of the final key after application of the function. This selection is beyond the scope of the current text, but the construction of these functions, even within the quantum domain is an abiding interest in the cryptographic community \cite{Zhang07, Ched10, Yang16}.

In the ideal scenario, $r_f=r$. This final step in distillation provides a layer of security, and so, should Eve know partial information of the random string $X_2$, and full information of $F$, she can still extract arbitrarily little information on the key $K=f(X_2)$ \cite{Benn88,Benn95}.

\section{Conclusion}

A state-independent quantum key distribution protocol with two-way classical communication (SIQKD2CC) is proposed that is a variation of BB84 in which the test for eavesdropping is conducted through the time-like Bell-CHSH inequality in light of the BTCV formalism \cite{Bruk04,Tayl04}. It is shown that this inequality form manifests naturally from the non-commutativity of observables, and does not necessarily require an assumption of non-classicality of the temporal correlations. The SIQKD2CC protocol includes a classical part for secure key distillation, which implements prominent features in current privacy amplification and information reconciliation techniques that rely on two-way classical communication and provably increase key-rate. 

The SIQKD2CC protocol provides a number of additional advantages over current key generation techniques apart from state-independence.  It provides raw key generation from temporal correlations that violate a Bell-type inequality. This protocol also retains qubits that would have otherwise been eliminated during checks of basis matching between Alice and Bob. It avoids a potential worst-case assumption in current QKD protocols: that Eve initially provides Alice and Bob the entangled qubit pairs, while addressing supraquantum attacks and higher dimensional attacks in the manner of \cite{Barr05,Acin06,Scar06,Shen17}. A detailed analysis of security in the light of monogamy/polygamy relations, and security against different attacks and weak measurements falls under the domain of future work.

\nonumsection{Acknowledgements}
\noindent
The author thanks Prof. V. Ravishankar, Soumik Adhikary, and Sooryansh Asthana of  the Indian Institute of Technology, Delhi, for insightful discussions. The author also thanks the Department of Science and Technology (DST), India, for the funding behind this work through the Women's Scientist Scheme [SR/WOS-A/PS-29/2013 (G)], and the Indian Institute of Technology, Delhi, for providing the resources to pursue this work.

%\bibliographystyle{plain} 
%\bibliographystyle{unsrt}       

%\bibliography{bibliography}

\end{document}